\documentclass[preprint,showpacs,preprintnumbers,amsmath,amssymb]{revtex4}

\usepackage{graphicx}% Include figure files
\usepackage{dcolumn}% Align table columns on decimal point
\usepackage{bm}% bold math

\begin{document}

\preprint{To appear in Phyisical Review Letters, in October}

\title{
Evidence for nucleosynthesis in supernova $\gamma$-process:
Universal scaling on p-nuclei
}

\author{T.~Hayakawa$^{1,3}$, N.~Iwamoto$^{2}$, T.~Shizuma$^1$, T.~Kajino$^{3,4}$, H.~Umeda$^4$, and K.~Nomoto$^4$}

\affiliation{
$^1$Advanced Photon Research Center, Japan Atomic Energy Research Institute, Kizu, Kyoto 619-0215, Japan.\\
$^2$Department of Nuclear Energy System, Japan Atomic Energy Research Institute, Tokai, Ibaraki 319-1195, Japan.\\
$^3$National Astronomical Observatory, Osawa, Mitaka, Tokyo 181-8588, Japan.\\
$^4$Department of Astronomy, School of Science, University of Tokyo, Tokyo 113-0033, Japan.\\}
\date{\today}% It is always \today, today,

\begin{abstract}
With analyzing the solar system abundance,
we have found two universal scaling laws concerning the p- and s-nuclei.
They indicate that the $\gamma$-process  in supernova (SN) explosions
is the most promising origin of the p-nuclei 
that has been discussed with many possible nuclear reactions and sites
for about fifty years.
In addition the scalings provide new concepts: an universality of the $\gamma$-process
and a new nuclear cosmochronometer.
We have carried out $\gamma$-process nucleosynthesis calculations
for typical core-collapse SN explosion models
and the results satisfy the observed scalings.
\end{abstract}

%Use showkeys class option if keyword
\maketitle

%%% Introduction

The solar system abundance ratio
is an important record of stellar nucleosynthesis and
the chemical evolution of the Galaxy.
Stable elements heavier than iron in the solar system 
are classified into three groups in the nuclear chart,
as 1) the main nuclei located near the $\beta$-stability line,
2) the isolated neutron-rich nuclei, and 3) the neutron-deficient nuclei.
The first and second groups are inferred to be synthesized
via two different neutron-capture reaction chains,
i.e. the slow neutron-capture process (s-process) along the $\beta$-stability line,
and the rapid neutron-capture process (r-process) in the neutron-rich side.
The solar system abundance shows specific indication that they have actually 
happened before the solar system formation.
The first evidence is the two abundance peaks near the neutron magic number 
corresponding to the s- and r-processes \cite{BBFH}.
The second evidence for the s-process is an empirical relation, $N{\cdot}{\sigma}$ $\sim$ constant,
where $N$ and ${\sigma}$ are the solar abundance and
the neutron capture cross-section \cite{Seeger65,Gallino98}.
The third nuclei are called the "p-nuclei",
which are located in the neutron-deficient side 
and have very rare isotope abundance ratios (about 0.1\% $\sim$ 1 \%).
The anti-correlation between the photodisintegration reaction rates and the solar abundances for the p-nuclei
has been pointed out \cite{Woosley78}.
Nevertheless, their origin has long been discussed
with many possible nuclear reactions, and 
their astrophysical sites have not been identified uniquely.
The proposed nuclear processes are the rapid proton capture reactions in novae and Type I X-ray bursts in
neutron stars (rp-process) \cite{Schatz98,Schatz01},
the proton-induced reactions by Galactic cosmic rays \cite{Audouze70},
the photodisintegration reactions in supernova (SN) explosions 
($\gamma$-process) \cite{Arnould76,Woosley78,Howard91,Fujimoto03},
and the neutrino-induced reactions in SN explosions 
($\nu$-process) \cite{Woosley90,Hoffman96,Goriely01}.
The origin of the p-nuclei is crucial to our understanding of how the solar-system material formed and evolved.
The first purpose of this article is to report
empirical laws obtained from a careful analysis of 
the solar system abundance, which
indicates that the origin of the p-nuclei is the SN $\gamma$-process.
The second purpose is to show our theoretical SN nucleosynthesis
calculations and to propose two new concepts: universality on p-nuclei 
in the SN $\gamma$-process and a nuclear cosmochronometer for the $\gamma$-process.

First we discuss the features of the p-nuclei in the solar system abundance.
There are twenty two p-nuclei associated with almost pure s-nuclei that 
have two more neutrons than the p-nuclei.
The pure s-nuclei are dominantly synthesized 
by the s-process and shielded by stable isobars
against the ${\beta}^-$-decay after the freezeout of the r-process.
A typical example is found in hafnium isotopes:
$^{174}$Hf is a pure p-nucleus and 
$^{176}$Hf is a pure s-nucleus shielded by $^{176}$Yb against the r-process,
while $^{175}$Hf between them is unstable.
Taking the abundance ratios 
of the s-nucleus to the p-nucleus,
N(s)/N(p), where N is each isotope abundance, we find a clear
correlation between them (see, Fig.~1).
The ratios concentrate at a constant value of
N(s)/N(p) $\approx$ 23 in a wide region of the atomic number.
Large deviations for Ce and W can be explained by an exceptional contribution
from the r-process because they are not shielded completely.
Small deviations for Cd, Sn and Gd originate from
a weak branch of the s-process which contaminates the p-nuclei.
Mo and Ru are known to have different origin 
from other p-nuclei in the previous studies \cite{Prantzos90,Rayet95}.
Deviations in the heavy mass region may originate from the large uncertainties of the solar abundances.
The uncertainties for $^{184}$Os and $^{190}$Pt are about 50 and 100\%, respectively\cite{Isotope}.
The measurement of their abundances with a high precision is desired.
Except for these deviations,
the scaling rule N(s)/N(p) $\approx$ constant 
holds for very wide region of the atomic number,
which has never been recognized in literature quantitatively.

Furthermore, we find another scaling rule
between two pure p-nuclei with the same atomic number.
Nine nuclear species have two pure p-nuclei,
in which the second p-nucleus is two neutron-deficient to the first
p-nucleus.
The abundance ratios, N(1st p)/N(2nd p),
are shown in Fig.~2.
Again the ratios are found to concentrate at almost constant value
with a weak slope as shown by the second scaling, N(1st p)/N(2nd p) $\approx$ 1,
to a very good approximation.
A large deviation for Er is attributed to a contamination from
${\beta}^-$-decay of $^{163}$Dy under stellar s-process conditions \cite{Takahashi83,Jung92}.

The first scaling shows a strong correlation between
p- and s-isotopes with the same atomic number indicating that the production mechanism
of the p-nuclei has the reason.
This is consistent with the previous theoretical calculations that 
the p-nuclei are produced by the $\gamma$-process
in SN explosions \cite{Prantzos90,Rayet95,Utsunomiya}:
namely, the pre-existing nuclei in massive stars are affected by the s-process
during the pre-supernova evolutionary stage
and the p-nuclei are subsequently produced from them by photodisintegration
reactions such as ($\gamma$,n) reactions
in a huge photon bath at extremely high temperatures in SN explosions. 
The calculations indicated that the p-nuclei are
produced via two paths, the direct ($\gamma$,n) reactions
and the EC/${\beta}^{+}$-decay from the neutron-deficient unstable nuclei 
which are first transmuted
by successive photodisintegration reactions ($\gamma$,n), ($\gamma$,p), ($\gamma$,$\alpha$) 
from heavier elements.
The first scaling suggests that the former reactions
are likely to play a role more important than the latter reactions.
On the other hand, 
the charged particle reactions in the rp-process \cite{Schatz98,Schatz01}
and proton-induced reactions by cosmic rays \cite{Audouze70}
change the proton number of seed nuclei.
In the $\nu$-process,
the charged current interaction that has a contribution larger than
the neutral current interaction also changes the proton number \cite{Goriely01}.
Therefore,
the scaling does not emerge from the dominant charged particle
processes or the $\nu$-process.
The first scaling is, thus,  a piece of evidence that the $\gamma$-process
is the most promising origin of the p-nuclei.

The N(s)/N(p) ratios
for the solar system abundance are subject to the Galactic chemical evolution.
The solar system formed from the admixed interstellar media 
originated from many different nucleosynthesis episodes in the Galaxy,
and the p-nuclei and s-nuclei were produced in different stellar
environments.
Thus, the mass distribution of synthesized nuclei may depend on the astrophysical
conditions.
Nevertheless, the observed N(s)/N(p) ratios in the solar system
are almost constant.
This leads to a novel concept that the N(s)/N(p) ratios produced by
individual $\gamma$-process are universal and
almost independent of the astrophysical conditions.

We carried out nucleosynthesis calculations of the $\gamma$-process
in oxygen-neon layers in SN explosions \cite{Iwamoto2003}.
The purpose of the calculations is to check the robustness
of our scaling  rules in the solar system abundance
and is to demonstrate the dependence of the calculated ratios, N(s)/N(p) and N(1st p)/N(2nd p),
on astrophysical conditions.
We used a solar metallicity ($Z=Z_\odot$), progenitor models
with 25 solar masses (25$~M_\odot$) which exploded with
an explosion energy of $10^{51}$ ergs.
The s-processed abundances for an initial chemical composition
were adopted.
The calculated N(s)/N(p) ratios are shown in Fig.~3 by open circles.
They show  almost constant abundance ratios in a wide region of the atomic number,
although the calculated N(s)/N(p) ratios show exceptional deviations.
This result is consistent with the observed scaling.
The observed ratios in the light mass region show a slight enhancement of the p-nuclei,
which may originate from progressively increasing roles of  ($\gamma$,p) and ($\gamma$,$\alpha$) reactions
with decreasing atomic number and/or the production from heavier nuclei at high temperature.
The calculated ratios are smaller than the observed ones by several factors
because the s-nuclei in the solar system mainly originate from the AGB stars \cite{Gallino98}.
In contrast, the relation, N(1st~p)/N(2nd~p) $\approx$ 1, can be directly compared with
the theoretical calculations of the SN $\gamma$-process, and thus
the second scaling rule can be used for constraining strongly the SN $\gamma$-process models.
The calculated N(1st~p)/N(2nd~p) ratios (open circles) in Fig.~2 are in reasonable
agreement with the empirical values (filled circles).

We further performed the $\gamma$-process nucleosynthesis calculations
for the 15 and 40 $~M_\odot$ progenitors
to study the progenitor mass dependence.
The abundance patterns of two ratios
do not change drastically from those in the 25 $~M_\odot$  models.
This result indicates that the two ratios are almost independent on the progenitor mass of the massive stars.
We calculated the $\gamma$-process in the different metallicity ($Z$=0.05~$Z_\odot$) models
with the same progenitor mass.
The calculated result show that the ratios are almost constant and
independent of the metallicity.
These results support  the proposed universality of the $\gamma$-process.
The calculated results in the previous $\gamma$-process studies were
shown to compare directly  with the solar system abundance 
of the p-nuclei \cite{Prantzos90,Rayet95,Utsunomiya}, 
not in the form of N(s)/N(p) or N(1st p)/N(2nd p) as we proposed 
in the article.
The $\gamma$-process calculations 
for different models constructed with different explosion energies 
or the $^{12}$C($\alpha$,$\gamma$)$^{16}$O reaction rate
showed the similar abundance distributions \cite{Prantzos90,Rayet95}.
These results also support the universality of the $\gamma$-process.
Although these results indicate that the p-nuclei in the solar system are mainly produced 
by the $\gamma$-process in Type II SNe,
other astrophysical sites such as deflagrating white dwarf stars \cite{Howard91}
and supernova-driven supercritical accretion disks \cite{Fujimoto03}
may also contribute to the p-nuclei.
Overproduction factors of p-process nuclei 
in realistic models of exploding stars were often a factor of a few below 
what is needed to explain the solar abundances.  
This may signal the $\gamma$-process in some
other environments as another producer of the p-nuclei.
We presume that such $\gamma$-processes should also reproduce the two scalings.

The universality of the SN $\gamma$-process as presented by the scalings 
is an important concept for understanding the
chemical evolution of the Galaxy.
The s-nuclei in the solar system were mainly produced
by the s-process in the low-mass AGB stars \cite{Gallino98}.
The  average N(s)/N(p) ratios are thus 
proportional to the abundance synthesized by individual s-process
and the frequency of the formation of the AGB stars.
The s-process nucleosynthesis depends highly
on the metallicity
which increases along the evolution of the Galaxy \cite{Raiteri92,Busso99}.
This fact concerning the s-nuclei and the universality of the p-nuclei
suggests that the N(s)/N(p) ratios may depend on time.
%%%
Astronomical observations of the time variation of these ratios for various metallicity stars
should constrain the Galactic chemical evolution of the s-nuclei
and also provide new information of the metallicity dependence of the s-process
nucleosynthesis \cite{Aoki2002}.
The recent progress in spectroscopic studies of extremely metal-poor stars has
enabled successfully isotope separation of several heavy elements  \cite{Lambert02,Aoki03b}.
It is of particular interest to observe the ancient metal-poor stars
whose material had been affected by the single or a few SN $\gamma$-processes.
Since the primitive gas is made of the products of the Big-Bang
nucleosynthesis or an explosive nucleosynthesis in the first generation
population-III SNe, it does not contain any heavy s-nuclei.
The abundance distribution of the p-nuclei in metal-poor stars
are, thus, expected to be very different from the solar abundance distribution
and the detection of p-nuclei by spectroscopically separating isotope abundances 
in these stars would be an urgent subject in
the future studies.

The universal scaling also plays a critical role in constructing
a chronometer that can be applied to analysis of pre-solar grains
in primitive meteorites which had been
affected strongly by a single or a few nucleosynthesis episodes.
Radioactive nuclei of cosmological significance are very rare and
only six chronometers with half-lives 
in the range of the cosmic age 1 $\sim$ 100 Gyr were known \cite{BBFH,Clayton64,Audouze72a}.
Historically, a new cosmochronometer with suitable half-life has not been proposed
for the last thirty years.
We here propose a new cosmochronometer
$^{176}$Lu (half-life 37.8 Gyr)-$^{176}$Hf-$^{174}$Hf
of the $\gamma$-process in the SN explosion.
Although the $^{146}$Sm and $^{92}$Nb have already been proposed as possible chronometers
of the $\gamma$-process \cite{Audouze72b,Harper96},
their half-lives are shorter than the age of the solar system.
Therefore, our proposed chronometer becomes a unique 
$\gamma$-process chronometer
which has a suitable time scale of order of the cosmic age  $\sim$ 10 Gyr.

A $^{176}$Lu - $^{176}$Hf pair was previously proposed as an s-process
chronometer \cite{Audouze72a}.
It was pointed out to be a good thermometer but useless as the chronometer \cite{Beer81}.
The initial abundance of the daughter nucleus, $^{176}$Hf, is uncertain
because the ${\beta}^-$-decay through the $^{176}$Lu isomer is accelerated 
at a typical s-process temperature.
Our proposed chronometer of the SN $\gamma$-process is completely free form this uncertainty.
The initial abundance of $^{176}$Hf
is calculated from the present abundance of $^{174}$Hf by applying 
the first scaling
if the pre-solar grain is affected strongly by a single SN event.
The first scaling indicates that the abundance of $^{174}$Hf is proportional
to $^{176}$Hf.
The passing time after the SN $\gamma$-process
can be calculated by,

\[
T =
-\frac{T_{1/2}(^{176}\mbox{Lu})}{ln2}{\times}
\]
\begin{equation}
ln\Biggl(\frac{N(^{176}\mbox{Lu
})}
{N(^{176}\mbox{Lu})+
\Bigl(N(^{176}\mbox{Hf})-R_i(\mbox{Hf}){\times}N(^{174}\mbox{Hf})\Bigr)}\
\Biggr),\label{chrono}
\end{equation}

\noindent
where, $N(^{A}Z)$ means the isotope abundance, and $R$ stands
for the N(s)/N(p) ratio in the scaling in meteorites,
which should be systematically measured 
or predicted by $\gamma$-process calculations.
Heavy elements such as Sr, Zr, Mo and Ba in primitive material
such as the pre-solar grains have
already been successfully
separated into isotopes including p-nuclei,
whose origin is considered to be the ejecta
of core collapse SN explosions \cite{Pellin00,Yin02}.
Although the pre-solar grains would be likely to condense $^{176}$Hf
and $^{176}$Lu from other regions of the star, the chemical composition of
the grains enhanced by the products of the O/Ne layer may be found.
The separation of three isotopes, $^{174,176}$Hf and $^{176}$Lu, 
in the pre-solar grains is highly desirable.

In summary, we presented two universal scaling laws concerning the
p- and s-nuclei in the solar system abundance.
They provide three novel concepts: a piece of  evidence that the SN $\gamma$-process 
is the most promising origin of the p-nuclei, 
an universality that the abundance ratios N(s)/N(p) of nuclei 
produced by individual SN $\gamma$-process are almost constant in
a wide region of the atomic number,
and a new nuclear cosmochronometer for the $\gamma$-process.
The scalings are useful for identifying the astrophysical sites of the p-nuclei
and limiting the contribution from other nuclear processes.
We carried out typical Type II SN $\gamma$-process calculations and the results
support the universality of the $\gamma$-process.
Therefore our proposals provide new insights into the chemical evolution
of the Galaxy as well as the SN $\gamma$-process.

\begin{acknowledgments}

This work has been supported in part by Grants-in-Aid for Scientific
Research (12047233, 13640313, 14540271, 15740168) and for Specially Promoted Research
(13002001) of the Ministry of Education, Culture, Sports, Science and Technology of
Japan. We would like to thank T. Tajima and  M. Fujiwara for valuable discussions and
encouragement.

\end{acknowledgments}

\begin{list}{}{}

\item[Figure 1]

Abundance ratios of pure s-nucleus to pure p-nucleus,
N(s)/N(p), in the solar system. The p-nucleus is two neutron-deficient
isotope from s-nucleus with the same atomic number $Z$.
The ratios are almost constant about 
N(s)/N(p) $\approx$ 23 in a wide region of the atomic number.
The inset displays the same quantities in the linear scale
except for $^{138}$Ce and $^{180}$W pairs which show large
deviations from the scaling value $\approx$ 23.
Deviations from the scaling and the uncertainty of the Os and Pt isotopes
are discussed in the text.

\item[Figure 2]

Abundance ratios of two pure p-nuclei, N(1st p)/N(2nd p). 
The first and second p-nuclei are respectively
two and four neutron-deficient isotopes from s-nucleus with the same atomic number $Z$.
The filled circles mean the observed ratios in the solar system.
The open circles stand for the calculated ratios.

\item[Figure 3]

Comparison of the calculated and observed abundance ratios,
N(s)/N(p). The filled and open circles mean the observed ratio in the solar system
and the calculated ratios.
The uppermost dotted line is N(s)/N(p) = 23. The dashed line displays the
average value of the calculated N(s)/N(p) ratios in the SN $\gamma$-process model,
and the two dot-dashed lines above and below this line are those multiplied by
factor 3 and 1/3, respectively.

\end{list}


\begin{thebibliography}{99}
\bibitem{BBFH}
E.M. Burbidge, G.R. Burbidge, W.A. Fowler,  F. Hoyle,
{\it Rev. Mod. Phys.} {\bf 29}, 548 (1957).

\bibitem{Seeger65}
P.A. Seeger, W.A. Fowler, D.D. Clayton,
Astrophys. J. {\bf 11}, 121 (1965).

\bibitem{Gallino98}
R. Gallino, {\it et al.},
Astrophys. J. {\bf 497}, 388 (1998).

\bibitem{Woosley78}
S.E. Woosley,  W.M. Howard, 
Astrophys. J. Suppl. {\bf 36}, 285 (1978).

\bibitem{Schatz98}
H. Schatz,  {\it et al.}, 
Phys. Rep. {\bf 294}, 167 (1998).

\bibitem{Schatz01}
H. Schatz,  {\it et al.}, 
Phys. Rev. Lett. {\bf 86}, 3471 (2001).

\bibitem{Audouze70}
J. Audouze,
Astron. Astrophys. {\bf 8}, 436 (1970).

\bibitem{Arnould76}
M. Arnould,
Astron. Astrophys. {\bf 46}, 117 (1976).

\bibitem{Howard91}
W.M. Howard, B.S. Meyer, S.E.Woosley, Astrophys. J. {\bf 373}, L5 (1991).

\bibitem{Fujimoto03}
S. Fujimoto,  {\it et al.},
Astrophys. J. {\bf 585}, 418 (2003).

\bibitem{Woosley90}
S.~E. Woosley, D.~H. Hartmann, R.~D. Hoffman,  W.~C. Haxton, 
Astrophys. J. {\bf 356}, 272 (1990).

\bibitem{Hoffman96}
R.D. Hoffman, S.E. Woosley, G.M. Fuller, B.S. Meyer, 
Astrophys. J. {\bf 460}, 478 (1996).

\bibitem{Goriely01}
S. Goriely, M. Arnould, I. Borzov, M. Rayet,
Astron. Astrophys. {\bf 375}, L35 (2001).

\bibitem{Isotope}
P. De Bievre and P.D.P. Taylor,  Int. J. Mass Spectrom. Ion Proc. 123, 149 (1993).

\bibitem{Prantzos90}
N. Prantzos, M. Hashimoto, M. Rayet, M. Arnould,
Astron. Astrophys. {\bf 238}, 455 (1990).

\bibitem{Rayet95}
M. Rayet, {\it et al.},
Astron. Astrophys. {\bf 298}, 517 (1995).

\bibitem{Takahashi83}
K. Takahashi, K. Yokoi, Nucl. Phys. {\bf A404}, 578 (1983).

\bibitem{Jung92}
M. Jung, {\it et al.}, Phys. Rev. Lett. {\bf 69}, 2164 (1992).

\bibitem{Utsunomiya}
H. Utsunomiya, {\it et al.},
Phys. Rev. C {\bf 67}, 015807 (2003).

\bibitem{Iwamoto2003}
N. Iwamoto, H. Umeda, K. Nomoto,  
International Symposium on Origin of Matter and Evolution of Galaxies, World Scientific, in press (2004).

\bibitem{Raiteri92}
C.M. Raiteri, R. Gallino, M. Busso,
Astrophys. J. {\bf 387}, 263 (1992).

\bibitem{Busso99}
M. Busso, R. Gallino, C.J. Wasserburg,
Ann. Rev. Astron. Astrophys. {\bf 37}, 239 (1999).

\bibitem{Aoki2002}
W. Aoki, {\it et al.}, Astrophys. J. {\bf 580}, 1149 (2002).

\bibitem{Lambert02}
D.L. Lambert, C.A. Prieto,
Mon. Not. R. Astron. Soc. {\bf 335}, 325 (2002).

\bibitem{Aoki03b}
W. Aoki, {\it et al.},
Astrophys. J. {\bf 592}, L67 (2003).

\bibitem{Clayton64}
D.D. Clayton, Astrophys. J. {\bf 139}, 637 (1964).

\bibitem{Audouze72a}
J. Audouze, W.A. Fowler, D.N. Schramm, Nature Phys. Sci. {\bf 238}, 8 (1972).

\bibitem{Audouze72b}
J. Audouze,  D.N. Schramm, Nature {\bf 237}, 447 (1972).

\bibitem{Harper96}
C.L.Jr. Harper, Astrophys. J. {\bf 466}, 437 (1996).

\bibitem{Beer81}
H. Beer,  F. K{\"a}ppeler, K. Wisshak, R.A. Ward, 
Astrophys. J. Suppl. {\bf 46}, 295 (1981).

\bibitem{Pellin00} 
M.J. Pellin,  {\it et al.}, Lunar and Planet. Sci. {\bf 31}, 1917 (2000).

\bibitem{Yin02}
Q. Yin, S.B. Jacobsen, K. Yamashita, Nature {\bf 415}, 881 (2002).

\end{thebibliography}
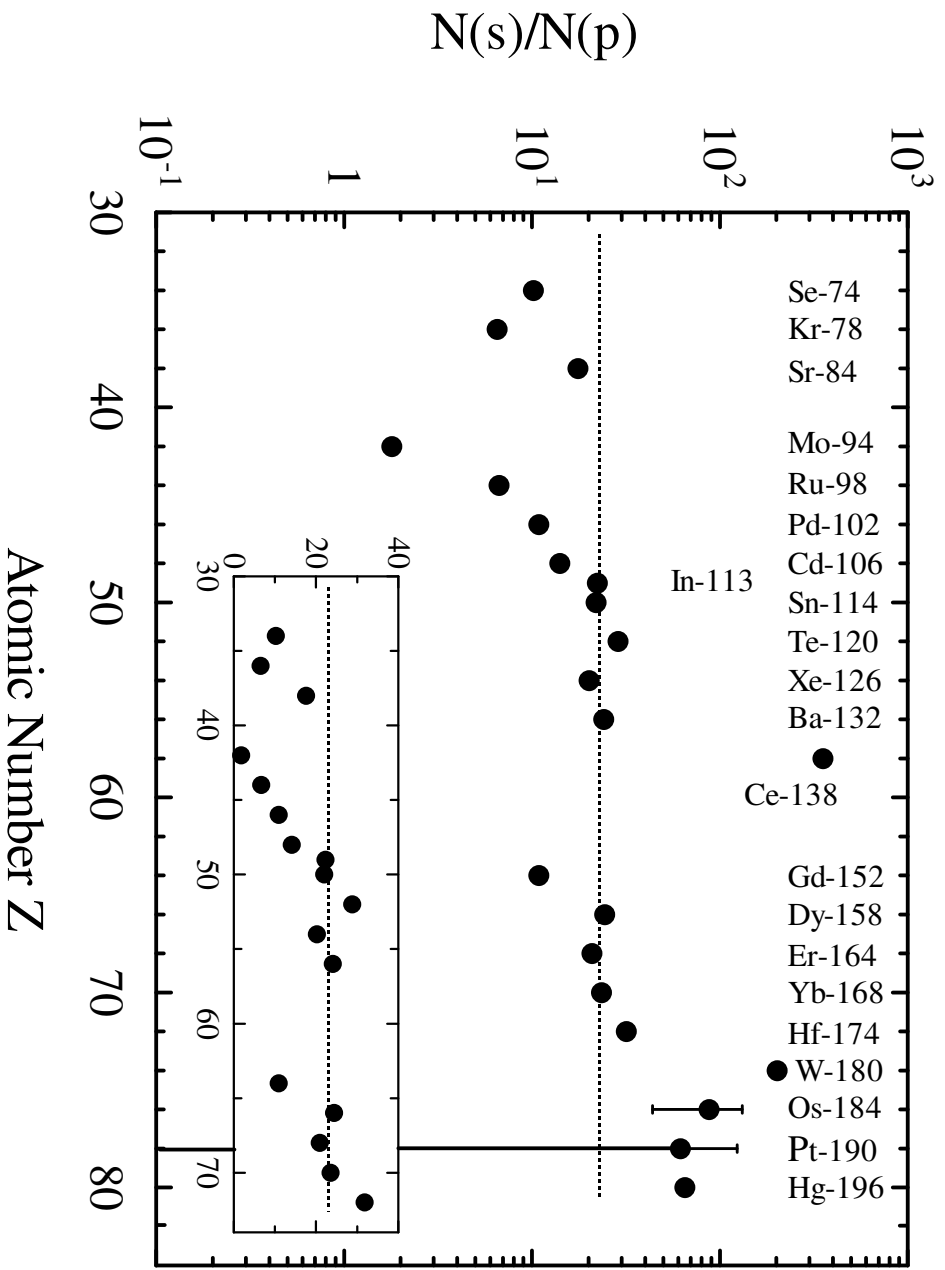
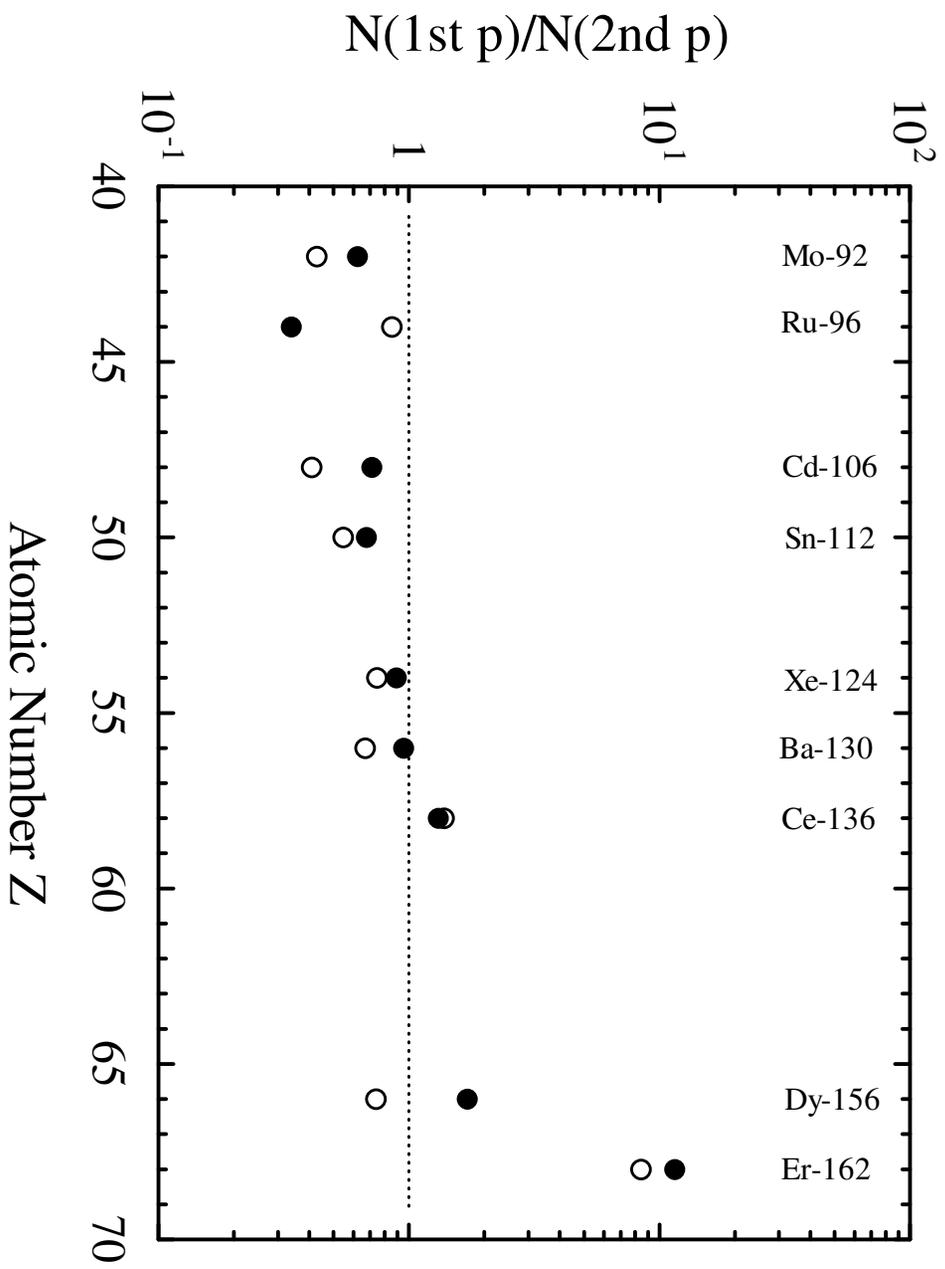
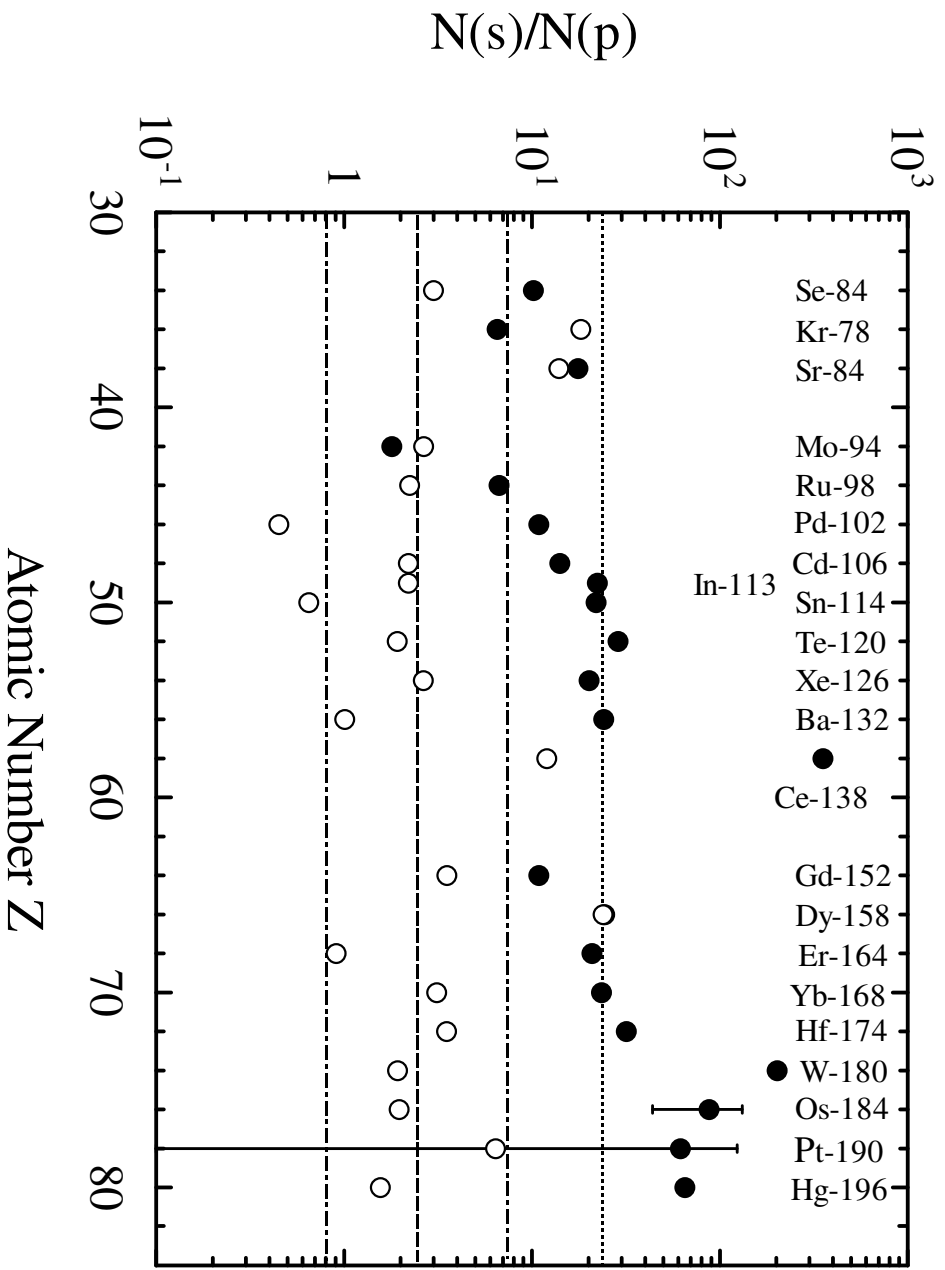
\end{document}